# A Model of Free Will for Artificial Entities

Eric Sanchis

University of Toulouse Capitole
Toulouse, France
e-mail: eric.sanchis@iut-rodez.fr

*Abstract*: **The impression of free will is the feeling according to which our choices are neither imposed from our inside nor from outside. It is the sense we are the ultimate cause of our acts. In direct opposition with the universal determinism, the existence of free will continues to be discussed. In this paper, free will is linked to a decisional mechanism: an agent is provided with free will if having performed a predictable choice C, it can immediately perform another choice C' in a random way. The intangible "feeling" of free will is replaced by a decision-making process including a predictable decision-making process immediately followed by an unpredictable decisional one.**

*Keywords*:    free will, artificial entities, agent, three-step decisionnal process, synthetic property.

## 1.  Introduction

Concepts produced by the human mind to characterize its own functioning, such as the properties of *intelligence*, *autonomy* or *free will* have a particularly important influence in our comprehension of the world. For this reason, they are continuously being revisited by philosophy, psychology, neurosciences, cognitive sciences but also computer sciences. In this latter area, it is undoubtedly the property of *intelligence* which has been the most studied in Artificial Intelligence, particularly in its most classical branch. Other properties, such as *autonomy* are studied in other fields of computer sciences (*Software Agent*, *Multi-Agent Systems* or *Robotics*), each one with their own issues.

More complex, *free will* is the property according to which the human being would be the first source of his choices. No internal or external force would impose him a choice rather than another. In other words, only the individual would be at the origin of his acts. After the first characterizations of this property, a question immediately emerged: does *free will* exist? This question has fuelled the philosophical debate for several centuries with no satisfactory solution found [1]. Many philosophers, such as Spinoza or Nietzsche, refuted its existence. Others, such as Sartre, did not feel that way. Others still have or have had a mitigated opinion. Studies of *free will* carried out by neuroscientists since the end of the last century, such as those achieved by B. Libet [2], although informative on certain points, did not give a definitive answer either.

Contrary to *intelligence* or *autonomy*, *free will* has received little attention in the computer science domain. A. Krausová and H. Hazan [3] examined the relevance to study *free will* within the *General Artificial Intelligence* field and answered by the affirmative. J. McCarthy [4] and R. Manzotti [5] tackled the *free will* problem within the classical Artificial Intelligence framework, the first via a logical formalization of the property reduced to a rational choice, the second by proposing a model sketch.

The model of *free will* which is proposed in this paper results from previous work related to the *autonomy* of software agents [6]. That is not completely surprising because these two properties have common characteristics, such as the plurality of possibilities and freedom of choice. *Free will* is considered as a decision-making process structured in three stages, where an agent makes a first choice that is immediately called into question by making a second choice randomly selected.

The paper is structured as follows. First, various aspects of *free will* are sketched, related to its nature and problems involved in the implementation of this property inside an artificial entity. The third section introduces a class of models called *Two-Stage Models of Free Will*. These models include the creation of the possible choices and the selection of the final choice. Sections 4 and 5 present the decision-making process at the heart of the *Three-Stage Model of Free Will*. In conclusion, entities being able to be provided with *free will* and the needed conditions are identified.

## 2. The Free Will Problem

*Free will* is a complex property which creates several questions. The first one is the following: how to characterize the property of free will in a precise way? In fact, there is no single definition! Indeed, depending on the language, certain formulations of the property name will privilege the aspect of choice (in French, *libre arbitre*), others the aspect of will (in English, *free will*). It would rather be a perception, a sense that our choices belong to us, that they are not imposed to us neither from the outside, nor by an interior force which we would not control. It is the feeling that we are the main cause of our acts. *Free will* is mentally defined by physical, emotional and intellectual sensations that a person feels, giving birth to a specific global "feeling". In the case of *free will*, this "sense" is expressed in terms of freedom, will and choice, a concise formulation of which could be: my will is free to choose.

The second question is: why *free will* is so important? It is possible to identify at least two main reasons:
- *Free will* is considered as a specific property to humankind. As our knowledge progresses on the animal societies, other properties we believed to be specific to human beings appear to be in fact … much less specific! Let us mention for example the capacity to use a complex language, to have empathy, to use a technical knowledge in tools construction.
- *Free will* constitutes a social pillar. Because we are provided with *free will* we are *morally responsible* for our acts and *criminally liable* before society. Moreover, various studies showed that people had a more social behaviour when they believed in the existence of *free will* [7]. Denying the existence of *free will* is undermining one of the human society bases.

The third question is certainly the most important: does *free will* actually exist? Serious arguments denying its existence were put forward by philosophers and scientists. The threat mainly comes from the *universal determinism* posed as the world's general functioning principle. This is why the contents of this principle must be examined before any study of the property of *free will*.

### 2.1 *Determinism* and *free will*

The concept of *determinism* is closely associated with the *principle of causality* (there cannot be an effect without a cause). *Determinism* is a *janusian* concept, i.e., a concept which has two faces like the Roman god Janus. Indeed, two kinds of determinisms can be distinguished: *concrete determinisms* and *speculative determinisms*.

*Concrete determinisms* are determinisms whose existence could be proven in the form of a natural law (e.g., the law of gravitation) or explicitly built (e.g., a computer program). *Speculative determinisms* are not supported by any proof of their existence. Only a body of facts and correlations suggest that they can exist. Examples of *speculative determinisms* are *social determinisms* and the most representative of them: the *universal determinism*.

The *universal determinism* is the principle according to which the succession of each event in the universe results from the *principle of causality*, the *past* and *natural laws*. According to the *universal determinism*, all is predetermined. That means that each event in the universe is given before it occurs. Thus, the existence of *free will* would be an illusion produced by our ignorance of the *past* and *natural laws*. On the other hand, the *universal determinism* is a fertile methodological framework because it is an essential engine of the scientific investigation.

As part of the *universal determinism*, it is important to distinguish the occurrence of an event from its predictability. Although an event is predetermined, its predictability may not be effective as it appears in chaotic systems. This absence of predictability is often justified as being due to a lack of information. This lack of information is also used to deny the existence of *indeterminism* or to interpret *chance*. As a result, a system can be *deterministic and predictable*, *deterministic and not predictable* or *not deterministic* (thus not predictable) if the *universal determinism* is refuted.

In short, the *universal determinism* (metaphysical principle) induces a methodology (scientific approach), which demonstrates the existence of *concrete determinisms* (natural laws). But it is also a fossilizing metaphysical principle, which denies that our experience as a person is the first source of our choices.

As neither the existence of the *universal determinism* nor the existence of *free will* can be demonstrated, philosophers have summarized the relationship between these two concepts with two points of view:
- *Incompatibilism*: the existence of a deterministic universe is in complete contradiction with the existence of *free will*. Between the two, it is necessary to choose.
- *Compatibilism*: there is no total opposition between a deterministic universe and *free will*. In particular, it is possible to freely act in a deterministic world.

### 2.2 The notion of *cause*

Using the *law of causality*, *determinism* invokes the concept of *cause*. But is this notion actually easy to handle?
The concept of *cause*
- Returns to the past, most of the time unknowable: when we have a thought *T* or we have carried out an action *A*, we cannot specify in an irrefutable manner what all their causes were.
- Is a multilevel concept: it applies at the same time at the social, psychological, biological and physical levels. We generally do not know how these various causal levels overlap.
- Is protean: in each context where it appears, it presents itself in a different form.

- Leads to a regression ad infinitum: the event *E* is the result of cause *C*, which is the result of cause *CC*, etc.

Like *determinism*, the notion of *cause* also presents two faces:
- It can be an enlightening explanation of a precise experiment. For example, when I drop a stone it falls (*effect*) because the stone is subject to the gravitation law (*cause*).
- It can be unknowable due to an unknown past and composite contents.

It results that in many cases, the general concept of *cause* brings more problems than solutions.

In conclusion, the model suggested to interpret and implement the property of *free will* within an artificial entity will replace the problematic concept of *cause* by the concrete concepts of *inputs* and *influences* while the couple *determinism/indeterminism* will be replaced by the concepts of *predictability/unpredictability*.

## 3. The Two-Stage Models of Free Will

The characterization of *free will* in the form of a process including two stages [8] is assigned to the American philosopher and psychologist W. James.
In this model of *free will*, a process is sequentially executed in two steps:
- At the first step, a certain number of possibilities are generated (some of them can be randomly created). Several futures are then possible: it is the *freedom* aspect of the agent which is expressed.
- In the second stage, one of these possibilities is chosen, choice in which the chance does not intervene any more: it is the *will* part of the agent (Figure 1).

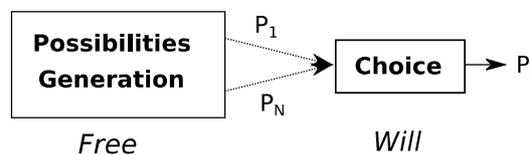

Figure 1: Two-Stage Model of Free Will

Since then, this model has been taken up directly or indirectly by several philosophers and scientists. B. Doyle listed a large number of these interpretations of *free will* that he classified under the generic expression *Two-Stage Models of Free Will* [9].

In Computer Science, J. McCarthy proposed a formalization of the concept which can also be considered as a *Two-Stage Model of Free Will*:
"*We present a theory of simple deterministic free will (SDFW) in a deterministic world. The theory splits the mechanism that determines action into two parts. The first part computes possible actions and their consequences. Then the second part decides which action is most preferable and does it*" [4].

The important common points to the different *Two-Stage Models of Free Will* are the following:
- The generation of the future possible choices and the creation of the selection function are parts of the *free will* process.
- There is only one phase of choice that ends the process associated with *free will*.

- In addition to the concept of *will*, these models use in the first stage properties difficult to characterize with precision, such as *intelligence* or *creativity*.

## 4. From the "Feeling" of Free Will to the "Mechanism" of Free Will

*Free will* is a complex property, i.e., it is difficult even impossible to have a thorough knowledge of it. It is a property with vague and elastic contours, with their contents changing according to the point of view that one can have. It authorizes different interpretations (ex: various *Two-stage Models of the Free Will*), when these interpretations are not contradictory. Lastly, for some *free will* exists, for others it does not exist.

Tackling *free will* by considering only the "feeling" associated to it or the "impression" it causes poses problems because it requires to use concepts which raise the same type of problems of definition and scope (ex: *conscience*, *will*, *freedom*, *first cause* of an act).

In order to solve these various problems, the model of *free will* described in the next section is based on the following points:
- *Conscience* and *free will* are decoupled. That means the implementation of *free will* in an entity does not require that this entity is provided with a brain or an advanced mental system.
- The *choice* aspect of *free will* is privileged over the *will* aspect. Indeed, the concept of *choice* can be defined with more precision than the concept of *will*.
- The *origin* and the *nature* of the causes related to a choice are ignored. The concept of *cause* is replaced by the concepts of *inputs* and *influences*.
- *Free will* is considered as a precise *decision-making process*. *Free will* is interpreted as the possibility to question a first choice by combining in the same decision-making process a *predictable choice* and a *random choice*.

## 5. A Three-Stage Model of Free Will

The suggested characterization of *free will* is the following: **an agent is provided with *free will* if, after having made a predictable choice *C*, it can immediately make another choice *C'* in a random way**. In other words, the impalpable "feeling" of *free will* is replaced by a *two-component decision-making process* including a *predictable decision-making process* immediately followed by a *unpredictable decision-making* one.

### 5.1 Model

The *Three-Stage Model of free will* implements two modules sequentially executed, a *predictable module* and a *unpredictable module*, driven by *causes* modelled in the form of *inputs* and *influences*.

¤ *Inputs and Influences*

We will say that a component (or module) has a *predictable behaviour* if the same *inputs* applied to it always generate the same *outputs*.

The term of *input* must be understood in its most general signification (internal or external conditions, stimuli, states, etc). It was chosen for simplicity reasons: the concept of *input* has the advantage of making the nature of the *causes* it summarizes transparent.

A module has an *unpredictable behaviour* if the same *inputs* applied to it at two different moments can produce different *outputs*.
When an *input* must be represented but with no or little information about it, the term of *influence* will be used. An *influence* materializes the complex aspect of a property by circumscribing its complexity in precise points of the model.

In the *Three-Stage Model of Free Will*, *influences* are only present in the *unpredictable component* (Figure 2).
Hereafter, the concept of *input* will be only used to represent well-identified information having an effect on a choice module. Otherwise, the term *influence* will be chosen.

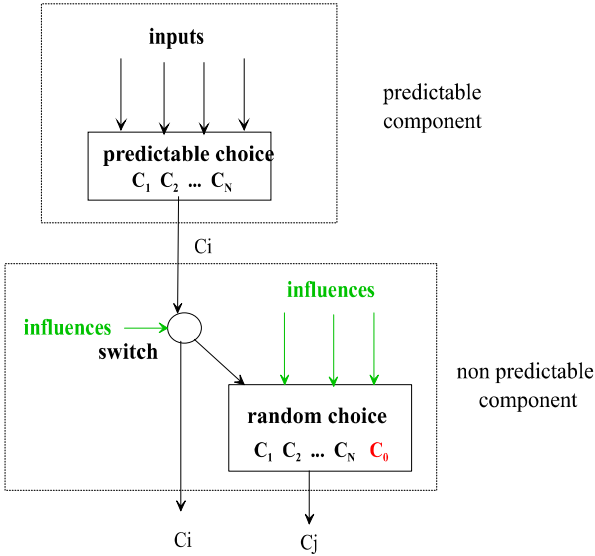

Figure 2: Three-Stage Model of Free Will

¤ *The predictable component*

The *predictable component* provides an artificial entity with a *regular behaviour*: this module can implement a *rational choice* (e.g., in situation *A*, my interest is to make choice *C*) or an *automatic behaviour* (e.g., each time I am in such a situation, I make this choice). In the first case, it is a thoughtful behaviour: the situation will be evaluated and the selected choice will be determined by taking into account various parameters, such as preferences, probabilities of occurrences of events, evaluation of consequences. In the last case, it is a constrained behaviour.
The presence of the *predictable module* also materializes the coherent behaviour of the agent: some sense pre-exists to the selected choice.

From a formal point of view, the *predictable module* is characterized by a *set of choices* $C_p$ ($1 \leq p \leq N, N > 1$), a *selection function* and a *set of inputs*. From the $N$ possible *choices*, the *selection function* and the *inputs*, the *predictable module* selects a choice $C_i$.
Contrary to the various *Two-Stage Models of Free Will*, processes involved in the development of these $N$ choices, the selection function and its inputs are considered being external to the property of *free will*. This means that it is necessary to have at one's disposal a set of choices, an associated selection function and the values of inputs before *free will* can appear.
The *predictable decision-making process* begins with the execution of the selection function. It is the first stage of the general process related to *free will*.

To illustrate the key elements of a *predictable module* we will use the following example. Let us suppose that one task of an agent *A* is to go to a point *P* no more than ten kilometres. A *regular* behaviour of *A* could be: if it rains agent *A* uses its car, if it does not rain and the sky is grey, *A* uses its bicycle. If the weather is nice, *A* reaches point *P* on foot.
The choice function of the agent is controlled by a unique input: the weather state. The three possible choices are: to go by car, to use its bicycle, to walk.

¤ *The unpredictable component*

This module includes two elements:
- A *switch*, which, either does not interfere on choice $C_i$ resulting from the *predictable component*, or activates an *unpredictable choice function*. In the first case, $C_i$ is the final choice. In the latter case, the final choice could be different.
- An *unpredictable choice function*, which carries out a random choice on the *N* choices available during the execution of the *predictable component* plus an *additional choice* noted $C_0$ called *empty choice*. This choice means that no choice is performed by the agent. It illustrates the situation where there is an inhibition of the choice resulting from the *predictable component*. Consequently, after the activation of the *unpredictable component*, the final choice $C_j$ ($0 \leq j \leq N$) may differ from choice $C_i$ selected by the *predictable component*. In particular, if $j=0$ there is inhibition of the choice resulting from the *predictable module*.

The following points must also be noted:
- There is no creation or development of new choices by the *unpredictable component*.
- Without context it is difficult to associate a precise meaning to the *empty choice $C_0$*. According to the situations, it can be interpreted in various manners like "*no choice must be made*" or "*it is a veto*".
- The activation details of the *switch* and *random choice function* can be partially or completely unknown to the agent. This is why their conditions of activation are represented in the form of *influences*.

Let us assume that in the previous example, the sun is shining. The *predictable component* invites agent *A* to reach point *P* on foot. But *A* can decide for reasons (*influences*) which are partly or completely unknown to it to call this choice into question: the *switch* is triggered. Then, *A* decides to select by random a choice among four: the three previous choices and the choice to remain where it is ($C_0$ choice). This *unpredictable choice* can be accomplished for example by using an appropriate computer program. It should be noted that the quality of the random choice function has a secondary importance in the global decision-making process.
In this precise example, the concept of *influence* makes it possible to represent an intuition as well as a change of mood of *A* having caused the switch triggering.

¤ *A Three-Stage decision-making process*

To sum up, the decision-making process associated to *free will* is structured by three moments (Figure 3):

- The moment $t_D$ when the execution of the *predictable choice function* begins. The *first stage* ends with the selection of a choice $C_i$.
- The moment $t_S$ when the entity decides to put into question the choice $C_i$. It is the *second step* of the decisional process.
- The moment $t_U$ when the execution of the *unpredictable choice function* begins, leading to the final choice $C$.

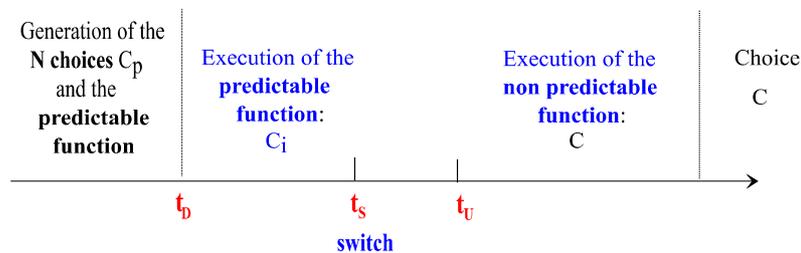

Figure 3: Temporal pattern of the decision-making process associated to free will

The agent has expressed its *free will* when the three moments $t_D$, $t_S$ and $t_U$ took place. This process will be qualitatively different according to the nature of the agent. If the agent is an individual, these three moments will be lived more or less consciously. For an artificial agent, the decision-making process associated to *free will* is simply executed.

## 5.2 Discussion

¤ *Model justification*

### a) Components articulation

Let us justify the role of the model elements and their articulation.
If the decision-making process associated to *free will* implemented only the *predictable choice module*, the agent executing this decision-making process would be comparable to a classical program like an accounting programme or a flight tickets booking program.
If the *unpredictable component* were the only one there, the agent would be stripped of rationality and condemned to an erratic functioning.
Let us suppose that the two modules of choice are reversed. Firstly, subjected to *influences,* the agent performs an *unpredictable choice*. In a second step, according to *influences*, the *switch* can give access to the *predictable choice module*, which according to the *inputs*, will select the rational choice. That means that the trigger of a coherent behaviour of the agent is controlled by *influences*, which is a very unsatisfactory functioning.
If there were no *switch*, the two modules would be sequentially executed. In this case, only the result of the last executed choice module would be considered. In one case, it boils down to always having a rational behaviour, in the other case to systematically exhibiting a random behaviour.
Lastly, the delicate concept of *ultimate cause* is replaced by the concrete notions of *input* and *influence*. Their values characterise the *current state* of the agent. The double advantage of the *current state* notion is that it avoids any reference to the origin of an event while synthesizing the agent's history.

### b) The feeling of free will

The choice to favour the *decision-making* aspect of *free will* rather than the *will* aspect makes it possible to design a model of this complex property. Conversely, the following question is worth asking: is this model compatible with the concept of *free will*? Formulated in another manner, the question becomes: does the proposed mechanism make it possible to recompose, at least partially, the feeling of *free will*?

From a human point of view, it is undeniable that this *three-stage decision-making process* can give an individual the feeling that he is provided with *free will* since he has the impression of being the source triggering the *unpredictable component*. However, that does not mean that it is the unique situation where an individual can have the feeling to be provided with *free will*.

¤ *Comparison between Two-Stage and Three-stage Models*

Several aspects deeply differentiate *Two-Stage* and *Three-Stage Models* (Figure 4).

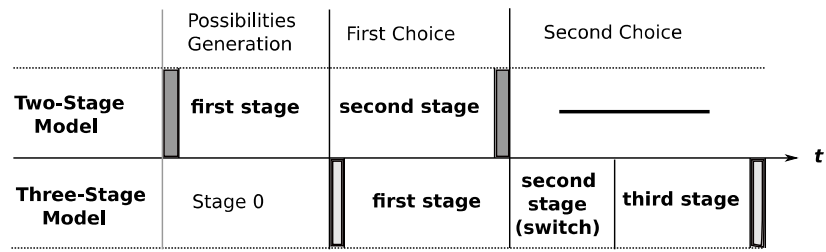

Figure 4: Two-Stage vs. Three-Stage Model

In the *Two-Stage Model*, the properties of *intelligence* and *creativity* intervene in the first phase of the model because the creation of the possible choices is part of the process related to *free will*. *Free will* can thus be seen as a composition of heterogeneous properties which are complex to represent. In the *Three-Stage Model*, choices creation is not part of the *free will* process. This possibilities generation step (*Stage 0*) takes place before the expression *of free will*. The result is a much finer granularity of the *Three-Stage Model* than that of the *Two-Stage Model*.

The *Two-Stage Model* only contains one phase of choice. This choice can be rational or not. It expresses the *will* aspect of the entity. The *Three-Stage Model* utilizes two consecutive choices, a predictable choice followed by a random choice. However, the concept of *choice* is much simpler to implement than the concept of *will*.

Lastly, the main weakness of the *Two-Stage Model* is not discriminating enough. Let us take as example a chess program. First, it will analyse the chess position, will generate the different possibilities and will choose the best move. The program functioning corresponds perfectly to the *Two-Stage Model of Free Will*. It is then legitimate to conclude that a chess program is provided with *Free Will*. According to the *Three-Stage Model*, a chess program is not provided with *Free Will*.

¤ *Model utilization*

*a) A critical analysis of a philosophical concept from a computational point of view*

The human cognitive system produces complex *mental concepts*, such as the concepts of *conscience*, *will*, *autonomy* and *free will*, partly characterized by other *mental concepts* with vague contents.

The computational approach of a concept, such as *free will* imposes a clarification of the essential components of the property to obtain an operational model of this concept. This clarification concerns at the same time the choice of the selected elements, their contents and their relationship.

Although this method leads to a simplified representation of the concept, it has the advantage of distinguishing with precision what was kept and what was left out of the property. When new knowledge is available or when the importance given to some features of the concept is changed, it will still be possible to update the model.

Conversely, reasoning about the model makes it possible on the one hand to identify, reduce and isolate the fuzzy areas of the property, and on the other hand to address its paradoxical aspects.

The result is a bidirectional questioning leading to a reciprocal enrichment between concept and model.

### b) Design of a synthetic property

The pursued approach is similar to that of synthetic biology: developing a *synthetic property* starting from data related to a precise philosophical concept. It is an engineering approach, which consists in improving the understanding of a property by disassembling and rebuilding it in a computational form after identifying its essential aspects.

It will be possible to put together this *synthetic property* with other *synthetic properties* in order to create *artificial characters*.

### c) Creation of artificial characters

The architecture of the *unpredictable component* provides a great flexibility to the creator of *synthetic characters*. Indeed, the concepts of *influence*, *switch* and *unpredictable choice* can be interpreted in various ways. This interpretative wealth allows the implementation of a broad spectrum of software agents.

For example, the agent's *unpredictable component* could be influenced by an *emotional module*. According to the composition of this module and its interconnection with the *unpredictable component*, various levels of steerability of the agent could be simulated. In this context, a second aspect, which could also be studied, is the degree of coupling between *emotions* and *free will*.

## 6. Conclusion

How to answer the question "Does *free will* exist?" It depends on the selected meaning of this property.

According to the meaning presented in this paper, *free will* exists because it is associated with a precise decision-making process. But as *free will* is a complex property with contradictory interpretations, it is not possible to give an absolute answer.

The *Three-Stage Model of Free Will* is a decision-making process relying on a mechanism that is possible to implement in an artificial entity: this model presupposes neither the existence of a brain nor a spirit. Consequently, a natural or artificial entity that is able to exhibit this decision-making process structured by these three moments will be considered as being provided with *free will*. According to this model, *free will* is a *global property* of the agent: either the agent possesses this property or it does not have it.

To the best of the author's knowledge, among the living beings, only humans are able to trigger this structured decision-making process and consequently are provided with *free will*.

# References

[1] R. Kane (ed.), The Oxford Handbook of Free Will, 2nd Ed. Oxford University Press, Oxford, 2011.

[2] B. Libet, C.A. Gleason, E.W. Wright, and D.K. Pearl, "Time of conscious intention to act in relation to onset of cerebral activity (readiness-potential). The unconscious initiation of a freely voluntary act", *Brain* (1983) 106: pp. 623- 642.

[3] A. Krausová and H. Hazan, "Creating Free Will in Artificial Intelligence". In Beyond AI: Artificial Golem Intelligence, pp 96-109, 2013.

[4] J. McCarthy, "Simple Deterministic Free Will", 2005. [Online] [retrieved: 12, 2017]. Available from:
http://www-formal.stanford.edu/jmc/freewill2/freewill2.html
http://informationphilosopher.com/solutions/scientists/mccarthy/SDFW.pdf

[5] R. Manzotti, "Machine free will: is free will a necessary ingredient of machine consciousness?". In: Hernández C. et al. (eds) From Brains to Systems. Advances in Experimental Medicine and Biology, vol 718 (2011), pp. 181-191. Springer, New York, NY. doi: 10.1007/978-1-4614-0164-3_15.

[6] E. Sanchis, "Autonomy with Regard to an Attribute", IEEE/WIC/ACM International Conference on Intelligent Agent Technology 2007 (IAT 2007), Silicon Valley, California (USA), pp. 39-42, 2007, doi:10.1109/IAT.2007.110.

[7] K. D. Vohs and J. W. Schooler, "The value of believing in free will: encouraging a belief in determinism increases cheating", Psychol Sci. 2008 Jan;19(1): pp. 49-54. doi: 10.1111/j.1467-9280.2008.02045.x.

[8] W. James, "The Dilemma of Determinism". In The Will to Believe and Other Essays in Popular Philosophy. Longmans Green and Co, New York, 1897.

[9] B. Doyle, Free Will: The Scandal in Philosophy. Information Philosopher, 2011. [Online] [retrieved: 12, 2017]. Available from: http://www.informationphilosopher.com/books/scandal/